# Spatially-Modulated Silicon Interface Energetics *via* Hydrogen Plasma-Assisted Atomic Layer Deposition of Ultrathin Alumina

*Alex Henning[§, *], Johannes D. Bartl[§, ‡, +], Lukas Wolz[§], Maximilian Christis[§], Felix Rauh[§], Michele Bissolo[§], Theresa Grünleitner[§], Johanna Eichhorn[§], Patrick Zeller[†, #, %], Matteo Amati[†], Luca Gregoratti[†], Jonathan J. Finley[§], Bernhard Rieger[‡], Martin Stutzmann[§], Ian D. Sharp[§, *]*

[§]Walter Schottky Institute and Physics Department, Technical University of Munich, 85748 Garching, Germany

[‡] Department of Chemistry, WACKER-Chair for Macromolecular Chemistry, Technische Universität München, Lichtenbergstraße 4, 85747 Garching bei München, Germany

[†] Elettra-Sincrotrone Trieste SCpA, SS14-Km163.5 in Area Science Park, 34149, Trieste, Italy.

*sharp@wsi.tum.de, *alex.henning@wsi.tum.de

KEYWORDS Atomic layer deposition; Aluminum oxide, Silicon surface charge density, Field-effect passivation

# ABSTRACT

Atomic layer deposition (ALD) is a key technique for the continued scaling of semiconductor devices, which increasingly relies on reproducible and scalable processes for interface manipulation of 3D structured surfaces on the atomic scale. While ALD allows the synthesis of conformal films at low temperature with utmost control over the thickness, atomically-defined closed coatings and surface modifications are still extremely difficult to achieve because of three-dimensional growth during nucleation. Here, we present a route towards sub-nanometer thin and continuous aluminum oxide ($AlO_x$) coatings on silicon (Si) substrates for the spatial control of the surface charge density and interface energetics. We use trimethylaluminum (TMA) in combination with remote hydrogen plasma instead of a gas-phase oxidant for the transformation of silicon oxide into alumina ($AlO_x$). During the initial ALD cycles, TMA reacts with the surface oxide ($SiO_2$) on silicon until there is a saturation of bindings sites, after which the oxygen from the underlying surface oxide is consumed, thereby transforming the silicon oxide into Si capped with $AlO_x$. Depending on the number of ALD cycles, the $SiO_2$ can be partially or fully transformed, which we exploit to create sub-nanometer thin and continuous $AlO_x$ layers deposited in selected regions defined by lithographic patterning. The resulting patterned surfaces are characterized by lateral $AlO_x/SiO_2$ interfaces possessing step heights as small as 0.3 nm and surface potential steps in excess of 0.4 V. In addition, the introduction of fixed negative charges of $9\times10^{12}$ $cm^{-2}$ enables modulation of the surface band bending, which is relevant to the field-effect passivation of Si and low-impedance charge transfer across contact interfaces. Overall, the described plasma-assisted ALD process allows the creation of ultimately thin microscopic patterns comprising laterally defined dielectric environments, chemical functionalities, and charge densities on Si substrates.

**INTRODUCTION**

Atomic layer deposition (ALD) has rapidly emerged as an essential tool in the semiconductor industry since it provides highly conformal and precisely tunable coatings with sub-nanometer thickness control at low temperature. As such, ALD is a powerful method for integration of dielectrics in advanced optoelectronics and has proved critical for the realization of emerging non-planar electronic devices.[1] In particular, amorphous aluminum oxide ($AlO_x$), which can be conformally grown by ALD on structured surfaces, is widely used in semiconductor technology for dielectric and chemical passivation,[2, 3] carrier-selective charge transfer across the interfaces of silicon (Si) solar cells,[4-6] and gate dielectrics in non-planar field-effect transistors,[7] as well as for diffusion barriers and protective coatings.[8, 9] When applied as a surface coating for field-effect passivation of Si, ALD $AlO_x$ introduces a high fixed negative charge density ($10^{12} - 10^{13}$ cm$^{-2}$),[10] which repels electrons from the surface and suppresses surface recombination *via* the field-effect, thereby improving the efficiency of Si solar cells.[2, 10, 11] Recently, ultrathin $AlO_x$ has also been implemented as an interlayer for hole-selective tunnel contacts to *p*-doped Si, which has been shown to further improve the performance of Si solar cells.[6, 12-14]

Chemical bonding directly at the $SiO_2/AlO_x$ interface predominantly governs the negative charge and interface state density, whereas the bond coordination within the dielectric film remains unchanged away from the interface.[10, 15] As a consequence, relatively thin ALD $AlO_x$ layers (several monolayers) can be used for field-effect passivation of *p*-doped Si, with such thin films exhibiting a similar impact as much thicker $AlO_x$ films.[3, 14, 16, 17] However, sub-nanometer thin and conformal ALD coatings are challenging to achieve due to precursor steric effects and non-uniform distributions of available binding sites on the sample surface, which lead to three-dimensional island growth during the nucleation regime.[3] Therefore, significantly lower fixed negative charge is introduced at the Si/SiO$_2$/AlO$_x$ interface for few-cycle ALD processes[14] since the deposited AlO$_x$ films tend to be discontinuous.[17] Beyond field effect passivation, the challenge of creating sub-nanometer thin *and* continuous ALD layers limits the ability to engineer semiconductor interfaces at this length scale, which is of critical importance for future device downscaling and low-impedance tunnel contacts.

In this work, we demonstrate an alternative approach to engineer Si surface energetics *via* ALD of ultrathin and conformal AlO$_x$ coatings formed during cyclic exposure to remote hydrogen (H$_2$) plasma and trimethylaluminum (TMA). Importantly, by using H$_2$ plasma instead of a gas-phase oxidant during ALD, AlO$_x$ formation is governed by the solid-state chemical reduction of the underlying SiO$_2$, resulting in an interconversion of the terminal oxide from SiO$_2$ to AlO$_x$. Here, *in situ* spectroscopic ellipsometry (SE) and quadrupole mass spectrometry (QMS) were used to elucidate the dominant growth mechanisms in both the few cycle limit and after extended ALD cycles, while X-ray photoelectron spectroscopy (XPS), atomic force microscopy (AFM), and surface photovoltage (SPV) measurements were used to characterize the AlO$_x$-terminated Si substrates. With the insights gained from *in situ* techniques, we demonstrate control of the AlO$_x$ thickness on a sub-nanometer level, resulting in significant modulation of the surface band bending through the introduction of fixed negative charge relevant to the field-effect passivation of Si. Moreover, the presented TMA/H$_2$ process can be performed at temperatures below 100 °C and, thus, is compatible with lithographic patterning. Taking advantage of this feature, we demonstrate smooth charge density patterns *via* spatially selective exposure of Si/SiO$_2$ surfaces to successive cycles of TMA/H$_2$ plasma. The resulting patterned surfaces are characterized by lateral AlO$_x$/SiO$_2$ interfaces possessing step heights as small as 0.3 nm, surface potential steps in excess of 0.4 V, and in-plane electric field strengths of ~3×10$^4$ V/cm. The ability to realize smooth dielectric and charge density patterns by low-temperature (LT) ALD in combination with lithographic techniques provides an intriguing possibility for creating carrier-selective contacts in Si solar cells and local gate structures in advanced transistors, as well as for spatially-defined surface functionalization[18] of relevance for area-selective deposition[19] and chemical sensing.[20, 21]

**RESULTS AND DISCUSSION**

For the case of native oxide-terminated III-V semiconductors, the application of TMA or of TMA/$H_2$ plasma cycles prior to deposition of dielectric coatings has been demonstrated to reduce the density of interface states.[22-26] These so-called oxide clean-up processes rely on ligand exchange between the metalorganic precursor and the oxidic surface, which is promoted by the strong electropositivity of TMA. However, to the best of our knowledge, analogous processes have not been investigated for the case of oxide-terminated Si, despite its prominent importance in semiconductor technology.

To gain mechanistic insights into the growth of $AlO_x$ on Si *via* successive exposure to TMA and remote $H_2$ plasma without the use of a gas-phase oxygen source, we tracked the ALD film thickness and reaction products in real-time by *in situ* spectroscopic ellipsometry and quadrupole mass spectrometry, respectively, and assessed the surface chemical composition by X-ray photoelectron spectroscopy. The evolution of the $AlO_x$ growth, derived from SE (**Fig. 1a**), reveals a steep decrease in the growth-per-cycle (GPC) during the initial ~10 cycles (at 200 °C), followed by a relatively slow decay of the growth rate during subsequent cycles. The decay in the growth rate throughout the deposition is well approximated with a bi-exponential function, suggesting two concurrent processes with distinct reaction rates. As such, we distinguish two $AlO_x$ growth regimes (denoted as (i) and (ii) in **Fig. 1a**), in which different surface chemical reactions dominate. The rapid decrease of the GPC during the initial cycles within regime (i) is consistent with a substrate-enhanced mechanism and the consumption of the available binding sites (*i.e.*, Si-O-H and Si-O groups) on the native $SiO_2$ terminated Si substrate,[27] while the formation of $AlO_x$ during growth regime (ii) is governed by diffusion-limited mass transport of oxygen from the underlying solid, as discussed in detail below.

Before film deposition, the Si substrate surface was cleaned in the growth chamber with a low-intensity remote $H_2$ plasma treatment (**Figs. 1b and S1**), resulting in a more hydrophilic (hydroxylated) surface (**Table S1**).[28] Subsequent exposure to TMA is known to lead to self-limiting adsorption of TMA according to the following reaction:[29, 30]

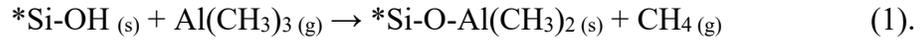

$$*\text{Si-OH}_{(s)} + \text{Al}(CH_3)_{3\,(g)} \rightarrow *\text{Si-O-Al}(CH_3)_{2\,(s)} + CH_{4\,(g)} \qquad (1).$$

Reaction 1 is one of several possible chemical routes for the adsorption of TMA on a hydroxylated surface.[29] For the case of traditional $AlO_x$ ALD processes, subsequent exposure of $*Si\text{-}O\text{-}Al(CH_3)_2$ to $H_2O$ leads to the regeneration of a hydroxyl-terminated surface in the form of *Al-OH, onto which TMA can bind in a subsequent half cycle.[29] However, in the present work, the surface is exposed to a remote $H_2$ plasma rather than a gas phase oxidant. Nevertheless, *in situ* QMS (**Fig. 1c**) indicates the evolution of $CH_4$, consistent with the following proposed sequential reaction within growth regime (i):[31]

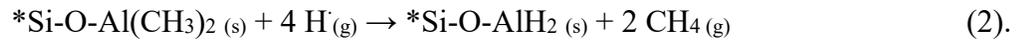

$$*\text{Si-O-Al}(CH_3)_{2\,(s)} + 4\ H^{\cdot}_{(g)} \rightarrow *\text{Si-O-AlH}_{2\,(s)} + 2\ CH_{4\,(g)} \qquad (2).$$

Upon additional cycling, the deposition rate within regime (i) decreases as Si-OH surface sites are consumed since the Al-terminated sites are considerably less reactive with TMA.[26] This proposed reaction pathway is supported by a comparative *in situ* SE experiment, in which the surface was exposed to an oxygen ($O_2$) plasma following 20 cycles of TMA/$H_2$ plasma (**Fig. S1**). Such a process results in an increase of the film thickness due to oxidation of Al-terminated surface sites. We note that this behavior is consistent with our prior work on GaN surfaces, in which a similar lack of TMA chemisorption after TMA/$H_2$ plasma exposure resulted in self-saturating deposition of an $AlO_x$ monolayer.[26]

Importantly, we find that the reactive hydrogen radicals generated by the $H_2$ plasma are essential for the release of the alkyl ligands of TMA *via* the generation of $CH_4$. To verify this, we performed a reference deposition using argon (Ar) plasma instead of $H_2$ plasma (**Figs. S1 and S2**). Despite the lack of reactive gas species, *in situ* SE reveals an increase of film thickness with cycle number (**Fig. S1**). Although the apparent alumina growth rate is larger for the Ar plasma process than for $H_2$ plasma process, subsequent oxygen ($O_2$) plasma exposure (**Fig. S1**) leads to a significant reduction of the film thickness for the TMA/Ar plasma process. Moreover, *in situ* QMS reveals a threefold reduction in the amount of methane released during the plasma step of the TMA/Ar process in comparison to the TMA/$H_2$ process (**Fig. S2**). These findings suggest that the TMA/Ar plasma process results in significant carbon incorporation due to physical decomposition of adsorbed TMA, with subsequent $O_2$ plasma exposure leading to volatilization of carbon impurities. In contrast, no such decrease in film thickness upon $O_2$ plasma exposure was observed for the $AlO_x$ film grown by the TMA/$H_2$ process (**Fig. S1**), indicating formation of $AlO_x$ with minimal carbon impurities, as confirmed by XPS (**Fig. S3a**). Thus, we conclude that Reaction 2 is a critical step for ligand removal during ALD alumina growth in the absence of a gas-phase oxidant.

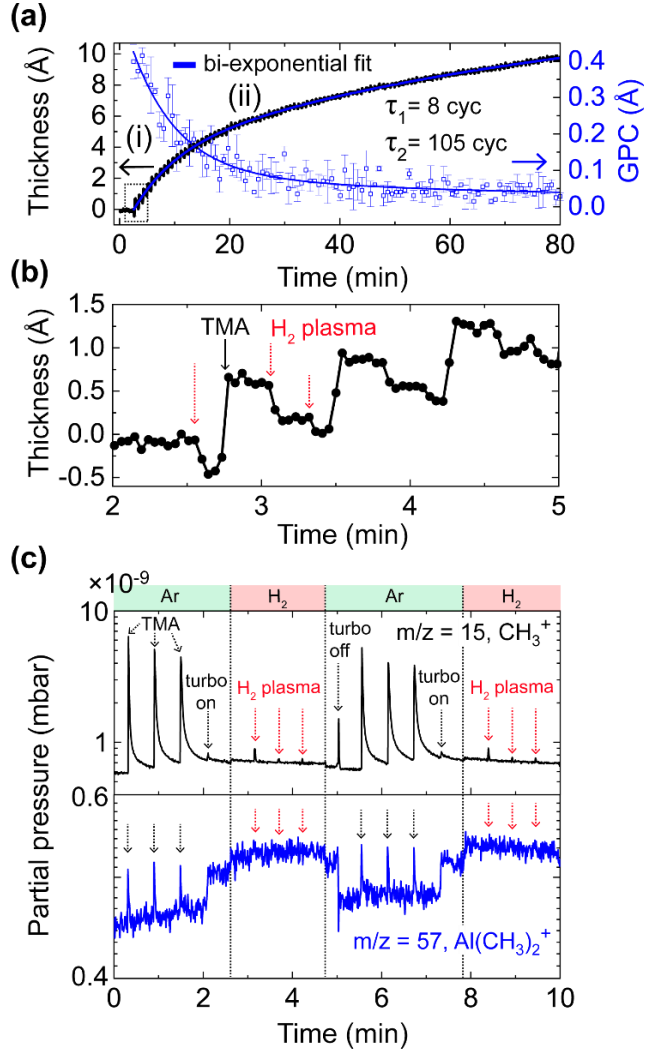

**Figure 1.** (a) Evolution of the AlO$_x$ thickness (black line) and the growth-per-cycle (GPC) (blue squares), derived from *in situ* spectroscopic ellipsometry, are shown for 100 cycles of consecutive exposure to TMA and remote H$_2$ plasma (100 W) at 200 °C. The decay in the GPC is approximated with a bi-exponential function that shows two concurrent mechanistic processes, indicated by regions (i) and (ii). (b) Zoom into the modelled thickness for three sequential cycles, indicating the impact of the TMA and H$_2$ exposure. (c) Tracking of the by-products during the introduction of TMA (3×) and H$_2$ plasma (3×) into the reactor chamber by *in situ* mass spectrometry. The CH$_3^+$ ions with a mass-to-charge (m/z) ratio of 15 stem from Al(CH$_3$)$_3$ (*i.e.*, TMA) or methane (CH$_4$), while Al(CH$_3$)$_2^+$ ions at m/z of 57 originate from TMA.

X-ray photoelectron spectroscopy provides additional insight into the nature of deposited AlO$_x$ films, as well as their mechanism of formation. Prior to ALD, the surface of the Si substrate is terminated by a native SiO$_2$ layer (**Fig. 2a**). In addition to the dominant silicon oxide component at a binding energy (BE) of ~103.3 eV, which is characteristic of $SiO_2$ ($Si^{4+}$), additional spectral contributions at lower BEs indicate the presence of sub-stoichiometric $SiO_x$, which has previously been associated with the $Si/SiO_2$ interface.[32] The XPS-derived thickness of the $SiO_2$ layer (~1.7 nm) agrees with the result obtained by SE (**Table S1**) and was calculated with the model proposed by Hill *et al.*[33] using the integrated intensities of the silicon oxide components, $I_{ox}$, and silicon component, $I_0$, of the Si 2p core level spectrum (**Fig. 2a** and **Supporting Information S3**). Following 20 cycles of TMA and remote $H_2$ plasma, the intensity ratio, $I_{ox}/I_0$, decreased (**Fig. 2b**), indicating a partial reduction of the surface $SiO_2$, which is consistent with the thermodynamically favored solid-state redox reaction of TMA with $SiO_2$ ($\Delta G = -235$ kcal/mol at 300 °C).[34] Indeed, the calculated $SiO_2$ layer thickness decreased by ~0.9 nm after the TMA/HP process. In contrast, application of conventional ALD of $AlO_x$ with TMA and $H_2O$ at 200 °C reveals an increase of the $SiO_2$ thickness of approximately 0.1 nm (**Fig. 2c**, **Table S1**). This result confirms the fundamentally different mechanism of $AlO_x$ growth using $H_2$ plasma compared to a gas phase oxidant ($H_2O$). In the former case, oxygen is supplied from the underlying $SiO_2$, resulting in its partial reduction.

In addition to the changing ratios of oxidized and elemental Si, the overall photoemission intensity from the Si 2p spectral region decreased following 20 TMA/$H_2$ plasma cycles due to attenuation by the overlayer (**Fig. S4**). Concomitantly, an additional O 1s component appeared (**Fig. S3b**) together with an Al 2p component at the characteristic BE of ~75.5 eV for Al-O[35] (inset of **Fig. 2b**). The resulting spectra are similar to those obtained from the conventional thermal TMA/$H_2O$ process, thus confirming the formation of alumina on Si after the TMA/$H_2$ plasma process. Here, we note that the sample was transferred from the ALD system to the XPS chamber through ambient air, which resulted in the oxidation of initially generated *Si-O-$AlH_2$ or *Si-O-Al($CH_3$)H sites.

Notably, the presence of AlO$_x$ alters the surface chemistry of Si, resulting in a more hydrophilic surface, as indicated by the decrease in the static water contact angle from 64° for the bare Si substrate to 25° for the TMA/H$_2$ plasma-treated Si (**Table S1**). In this way, the oxide conversion process can be exploited for subsequent (area-selective) chemical functionalization of the AlO$_x$-terminated surface, as recently reported by our group.[26, 36, 37]

We next turn our attention to growth regime (ii) (**Fig. 1a**), during which film deposition proceeded at a slow but nearly constant rate (~0.05 Å/cycle), despite the absence of a gas-phase oxygen source. Importantly, X-ray photoelectron spectra from samples exposed to 100 cycles of the TMA/H$_2$ plasma process, which is well within regime (ii), indicate that the native SiO$_2$ layer on Si was fully reduced to elemental Si, leaving a Si substrate coated with AlO$_x$ (**Fig. 2d**). We note that other constituents, such as AlSi$_x$O$_y$ may be present,[15] but were not detectable with the XPS system used in this study. Such complete transformation of the ~17 Å thick SiO$_2$ layer (**Fig. 1a**), composed of several monolayers, implies the solid-state transport of atomic or molecular oxygen through the SiO$_2$ and AlO$_x$ layers to the surface during the AlO$_x$ growth process. This is confirmed by XPS analysis of a quartz substrate after TMA and H$_2$ plasma treatment, which reveals the presence of Si$^0$ (**Fig. S5**) and thus confirms the complete reduction of SiO$_2$ to elemental Si. The activation energy for the diffusion of molecular oxygen through SiO$_2$ has been predicted by density functional theory to be 0.3 eV,[38] which is lower than the reported activation energy of aluminum diffusion in amorphous SiO$_2$ of 0.73 eV.[39] Although the thermal energy at the growth temperature of 200 °C ($k_BT$ = 0.041 eV) is significantly below the diffusion activation energies of both elements, it is far more likely for oxygen than for Al to diffuse within SiO$_2$.

In addition to the above considerations, a small shoulder indicative of aluminum carbide (AlC$_x$)[40] appears at low binding energies of the Al 2p core level region following 100 cycles of TMA/H$_2$ plasma (**Fig. 2d**, inset). Based on these observations, we infer that oxygen extraction is facilitated by a strongly oxidizing surface layer (adsorbed TMA) and that complete depletion of the solid-state oxygen source after extended cycling leads to increased carbon contamination in the form of aluminum carbide (AlC$_x$) due to the absence of an external oxidant. Thus, from the combined insights gained by QMS, SE, and XPS, we conclude that hydrogen plasma and TMA chemically react with the oxidized Si surface until there is a saturation of bindings sites (regime (i)) and subsequently until complete depletion of the oxygen (regime (ii)) from the native silicon oxide layer, thereby transforming the SiO$_2$ into Si capped with AlO$_x$.

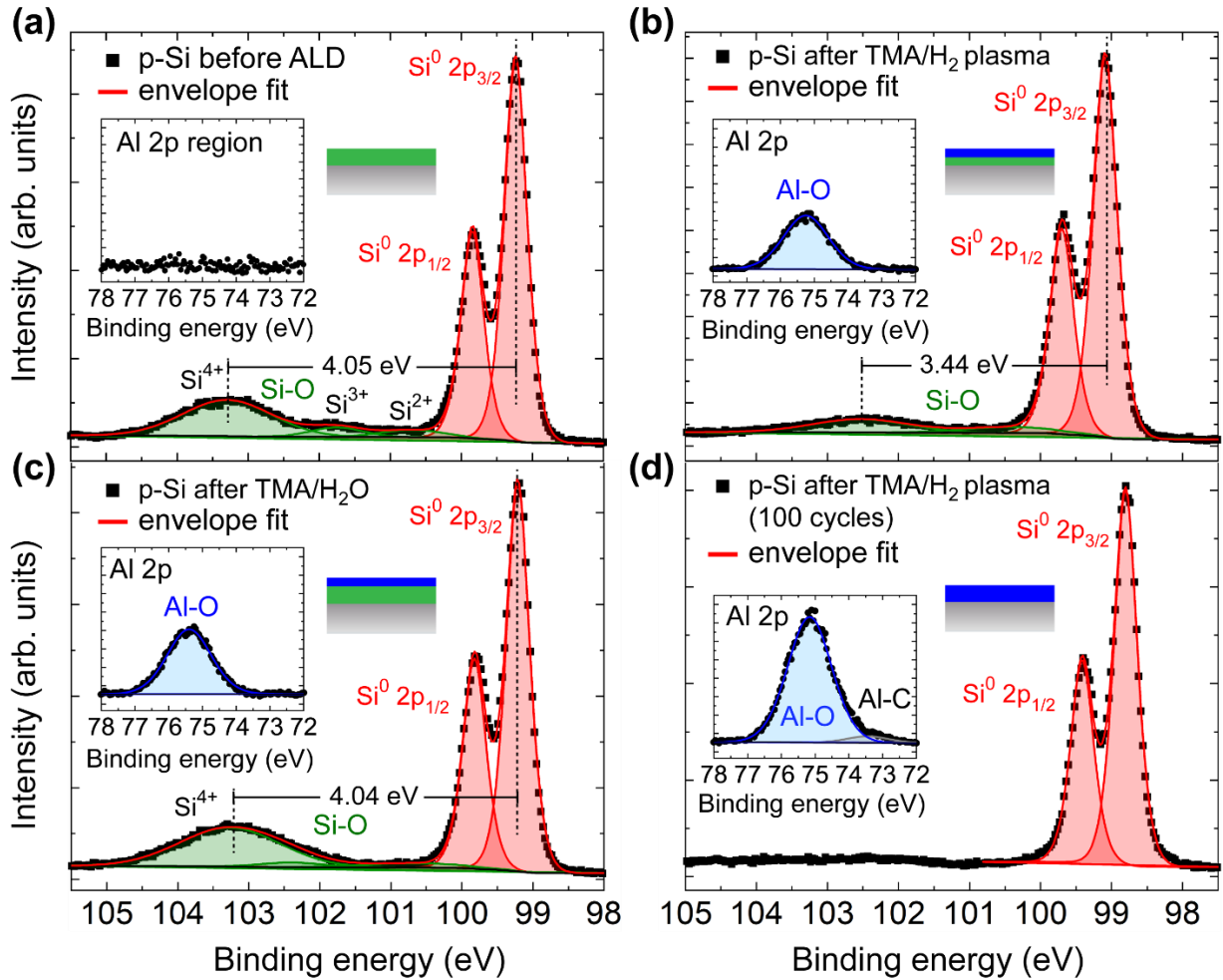

**Figure 2.** X-ray photoelectron spectra reveal the formation of aluminum oxide on a Si substrate following the transformation of the native $SiO_2$ layer after the $TMA/H_2$ plasma process. Si 2p core level spectrum of a Si substrate (a) with native $SiO_2$ before ALD, (b) after 20 cycles of TMA and remote $H_2$ plasma, (c) following ALD with TMA and water, and (d) after 100 cycles of TMA and remote $H_2$ plasma. The ratio between elemental Si ($Si^0$) and $SiO_2$ ($Si^{4+}$) decreased after the $TMA/H_2$ plasma process, while it slightly increased after the $TMA/H_2O$ process. The insets show the Al 2p core level region with the same intensity scale (*y*-axis) for comparison. The schematic cross-sections in the insets illustrate the Si substrate (grey) coated with silicon oxide (green), alumina (blue), or both oxide layers.

While the surface oxide is reduced for the $TMA/H_2$ plasma process, it is oxidized for standard ALD $AlO_x$ processes using water ($H_2O$) or oxygen plasma,[17, 41] the latter of which generates defects and introduces detrimental interface states in Si.[42] Moreover, for low-temperature ALD processes involving water as the oxidant source, condensation hampers self-limiting deposition and can reduce the film quality. Owing to the high vapor pressure and reactivity of TMA, as well as the self-limiting nature of the reactions of TMA and hydrogen radicals with the sample surface (Reactions 1 and 2), the ALD process can be performed at temperatures below 100 °C and is thus compatible with lithographic processing. Here, we exploited this feature to create dielectric micropatterns comprising distinct regions terminated with $SiO_2$ and $AlO_x$.

Scanning electron microscopy (SEM) and scanning photoelectron microscopy (SPEM) depict the $AlO_x$ pattern on a Si substrate (**Figs. 3 a,b**). The SEM image (**Fig. 3a**) shows a strong contrast between the two materials, $AlO_x$ and $SiO_2$, and confirms lithographic pattern fidelity after the lift-off process. Such a strong contrast in the SEM image, despite a sub-nanometer step height, can be explained with Dionne's model that predicts a larger secondary electron yield for ALD alumina than for $SiO_2$ (and Si) because of the higher material density, as well as the lower conductivity and electron affinity, of the former.[43, 44] From the elemental map of the Al 2p region obtained by SPEM (**Fig. 3b**) it is evident that alumina is deposited in spatially selected regions as no alumina signal (Al 2p) was detected within the $Si/SiO_2$ region.

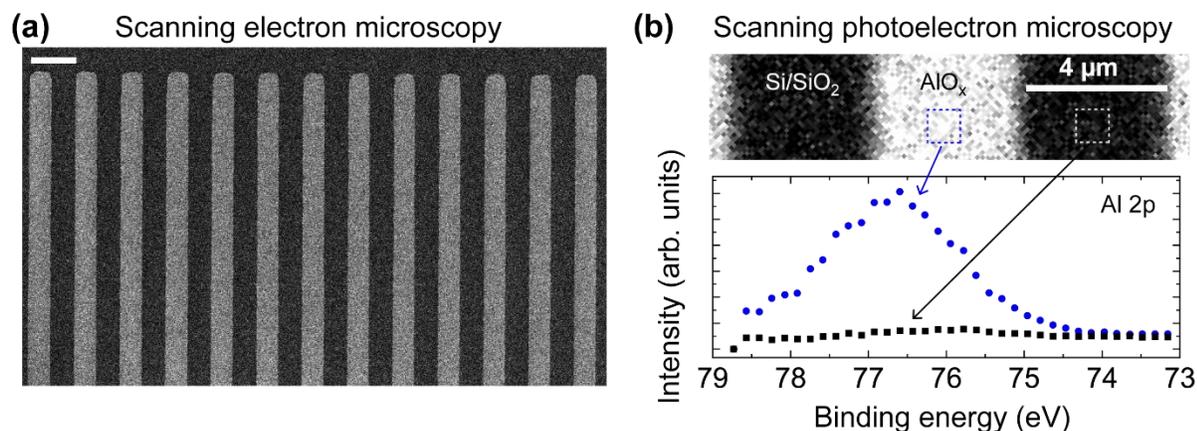

**Figure 3**. (a) Secondary electron scanning electron microscopy image of an $AlO_x$ pattern on Si substrate following 20 cycles of TMA and $H_2$ plasma at 70 °C. The bright regions correspond to $AlO_x$-terminated surfaces. The scale bar corresponds to 8 µm length. An electron acceleration voltage of 2.0 kV was used. (b) Synchrotron-based scanning photoelectron microscopy map of the Al 2p core level. Each pixel (150×150 $nm^2$) represents a spectrum in the Al 2p core level spectral range of BEs from 79 – 73 eV. The sample was excited with photons of 638 eV energy. The photoelectron core level spectra (bottom) were retrieved from the marked square regions in the SPEM image. We note that the spectrum is shifted towards higher binding energy due to differential charging of the insulating $AlO_x$ layer.

To assess the uniformity and thickness of the $AlO_x$ coating on a nanometer-scale, we characterized the surface morphology and potential landscape of the $AlO_x$ patterns by atomic force microscopy (AFM) and Kelvin probe force microscopy (KPFM), respectively (**Fig. 4**). Both the topography and the simultaneously acquired contact potential difference (CPD) maps are uniform within each region, demonstrating a continuous $AlO_x$ coating. The profile extracted from the AFM topograph (**Fig. 4c**, top) shows a height difference between the uncoated and $AlO_x$-covered Si region of only $3 \pm 0.3$ Å, which is thinner than previously demonstrated for conformal ALD $AlO_x$ coatings on Si. Notably, both surfaces are smooth, exhibiting a root mean square (rms) roughness ($R_{rms}$) of 0.18 nm (**Figs. 4 a, c**). The fact that the surface remains smooth after the TMA/$H_2$ plasma process is an essential criterion for its implementation in Si semiconductor technology.

Importantly, application of the TMA/$H_2$ plasma process for the formation of a sub-nanometer thin $AlO_x$ coating on Si substantially reduces its work function, as determined by KPFM. As such, this process provides opportunities to tune the energy band alignment at the Si interfaces, as discussed below. In particular, a similar decrease in the CPD of 0.4 V was measured for both *p*- and *n*-doped Si substrates following deposition of the AlO$_x$ layer of different thicknesses (**Figs. 4** and **S6**) and can be explained by the formation of a microscopic surface dipole, whose negative charges at the SiO$_2$/AlO$_x$ interface[10] are compensated by the positive charges from the AlO$_x$-terminated surface in ambient. Thus, a microscopic dipole layer with its positive pole pointing outwards leads to a decrease in the effective electron affinity and the observed reduction of the CPD and work function of the AlO$_x$ coated Si. Moreover, the potential profile across the patterned region reveals an abrupt potential step of ~0.42 V, yielding an electric field of ~3×10$^4$ V/cm at the lateral interface between bare and AlO$_x$-coated *p*-doped Si (**Fig. 4c**, bottom) whose magnitude is determined by the photolithography process and is experimentally underestimated due to the finite tip radius of the platinum-iridium-coated Si tip of about ~30 nm. Notably, the CPD is maintained over several months in ambient air as the same potential step has been repeatedly measured throughout this time.

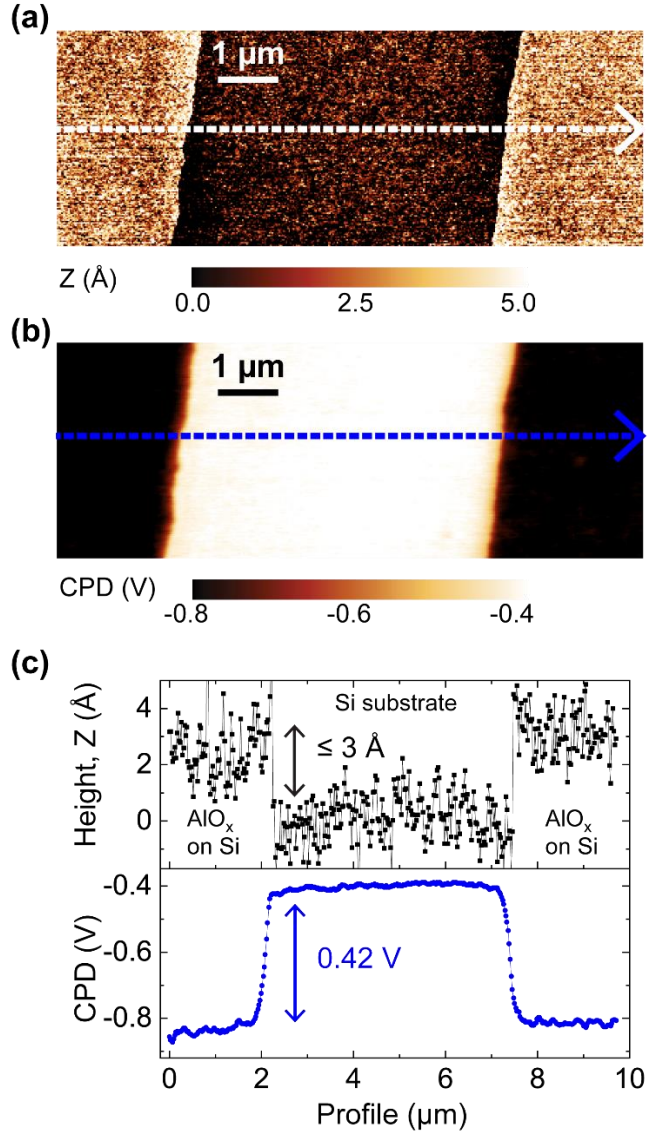

**Figure 4.** (a) Intermittent contact mode AFM topography and (b) frequency modulation KPFM image of the AlO$_x$ pattern on *p*-doped Si substrate. (c) Topography and CPD profiles were taken along the lines, as indicated in the images of (a) and (b). While a step height of less than 3 Å exists across the lateral interface between AlO$_x$ and SiO$_2$ terminal surfaces, a surface potential step of 0.42 V arises due to fixed charge introduced by the TMA/H$_2$ ALD process (10 cycles).

The observed reduction of the work function after alumina deposition on Si and the introduction of negative fixed charges is consistent with prior investigations of conventional ALD using TMA and H$_2$O, which has shown that AlO$_x$ deposited on Si introduces negative fixed charge that is beneficial for the field-effect passivation of $p$-doped Si.[2] Therefore, we investigated how the ultrathin AlO$_x$ coatings, synthesized *via* the TMA/H$_2$ process, affect the surface band bending (SBB) of moderately doped Si substrates of relevance for Si photovoltaics. Surface photovoltage measurements were conducted using a Kelvin probe upon above bandgap illumination with an 880 nm light emitting diode. As shown in **Fig. 5a**, these measurements reveal a significant increase (decrease) of the surface band bending of a moderately $n$-doped ($p$-doped) Si substrate after 20 cycles of the TMA/H$_2$ process. These observations confirm the presence of considerable negative charge density introduced by the AlO$_x$. For the case of $p$-type material, holes are attracted to the surface by the fixed negative oxide charge, resulting in an accumulation layer and the complete elimination of a surface photovoltage (**Fig. 5b**). Thus, ultrathin AlO$_x$ formed by the TMA/H$_2$ process can be used for field-effect passivation of $p$-doped Si. In contrast, the fixed negative oxide charge repels electrons within $n$-type material, resulting in a significantly larger surface band bending (**Fig. 5b**), which increased by 340 meV. Applying Poisson's equation and Gauss' theorem for the one-dimensional model of $n$-type Si shown in **Fig. 5b** with the measured surface band bending (**Supporting Information S5**), yields an increase in the fixed negative charge density at the AlO$_x$/SiO$_2$ interface of $Q_{it} = 9 \times 10^{12}$ cm$^{-2}$. We note that a similar increase of the SPV and surface band bending (350 mV) is achieved for $n$-type Si coated with a 1 nm thick AlO$_x$ layer deposited with the TMA/H$_2$O process (**Fig. S7**). This finding suggests that remote H$_2$ plasma exposure (and reduction of the SiO$_2$ layer) during the TMA/H$_2$ process does not introduce additional electronic defects and that the negative fixed charges are indeed introduced at the SiO$_2$/AlO$_x$ interface, in agreement with prior work.[3, 10, 14, 15] Although nanometer-thin alumina films, deposited by TMA and water, provide similar field effect passivation as the ones created with the TMA/H$_2$ process, these thicker layers impede interfacial charge carrier transport[45] and would lead to larger lateral step heights when patterned.

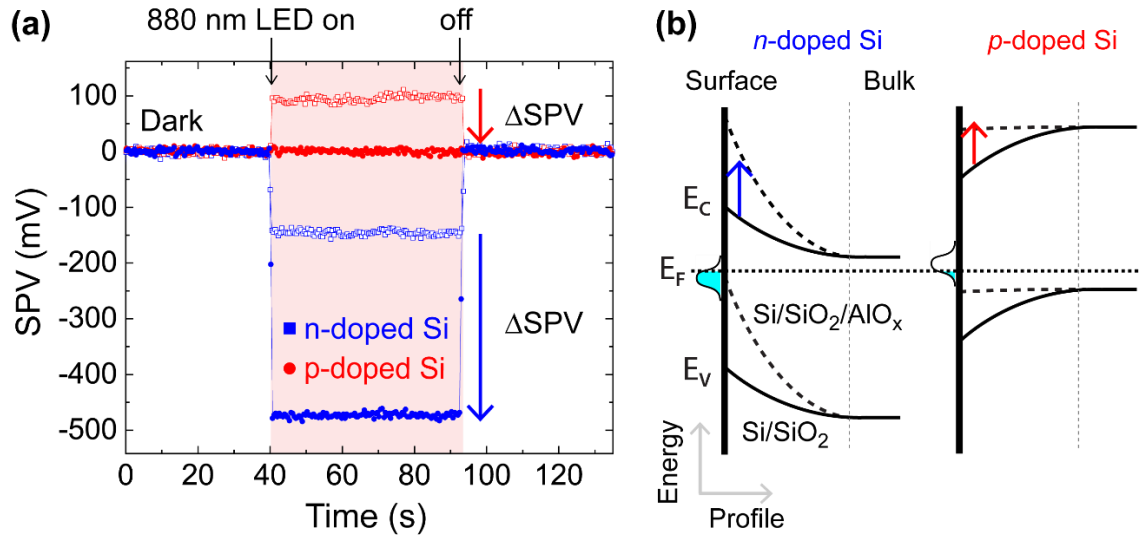

**Figure 5**. (a) Plots of the surface photovoltage as a function of time upon switching the 880 nm illumination source on and off for *p*-doped Si (red) and *n*-doped Si (blue) before (open circles) and after (filled circles) processing by 20 cycles of TMA and remote $H_2$ plasma. The vertical arrows indicated the magnitude of SPV change induced by the 20 cycle TMA/$H_2$ plasma treatment for both *n*- and *p*-type Si. The SPV is determined by subtracting the CPD values measured under illumination and in the dark, SPV = $CPD_{light}$ − $CPD_{dark}$. (b) Schematic energy band diagrams of *n*-doped Si (left) and *p*-doped Si (red).

**CONCLUSION**

In summary, we presented an ALD process based on TMA and remote hydrogen plasma to convert the native oxide on Si into alumina. This gas-phase process is scalable and omits wet-chemical treatments, thus providing a simple path to interface engineering in Si electronics. In particular, the sub-nanometer thin continuous $AlO_x$ coatings on silicon were demonstrated for field-effect passivation and can be potentially utilized for carrier-selective transport across contact interfaces. Furthermore, the selective deposition in lithographically defined areas at low temperatures allowed us to locally modify the dielectric environment and surface chemistry of Si substrates. The presented gas-phase processing concept provides new opportunities for engineered interfaces of relevance for surface functionalization of (bio)chemical sensors, control of charge carrier transport and field-effect passivation at Si interfaces, and defining gate structures for local control of charge carriers *via* field-effect. In addition, the ultrathin $AlO_x$ coating can be potentially used as a buffer layer to relax lattice matching constraints in the synthesis of semiconductor heterostructures, as previously shown for gallium nitride.[46]

## MATERIALS AND METHODS

**Silicon substrates.** Moderately *n*-doped (As) and *p*-doped (B) Si substrates (Siegert Wafer, Germany) with a <100> crystal orientation and a resistivity of ~6 Ω cm were used for this study. The frontside surfaces were chemo-mechanically polished by the manufacturer and were terminated with a 1-2 nm thick native oxide layer. The thicknesses of the oxide layers were determined by spectroscopic ellipsometry (SE) and X-ray photoelectron spectroscopy (XPS).

Ohmic contacts were applied to the backsides of the Si wafers by metal evaporation and annealing. Immediately prior to mounting samples in the evaporator, Si backside surfaces were etched in 5% HF solution to remove the oxide and provide passive H-terminated surfaces. All metallization procedures were carried out using thermal evaporation at rates between 1 and 3 Å/s under vacuum (<$10^{-6}$ mbar). For *n*-type Si, 100 nm of Al was evaporated onto the backside of the wafer before annealing in vacuum (~$10^{-6}$ mbar) at 500 °C for 5 min in a rapid thermal annealer. For *p*-type Si, 100 nm of a $Au_{0.99}Sb_{0.01}$ alloy was evaporated and annealed under the same conditions, resulting in the formation of a AuSi eutectic. Following Ohmic contact formation, 10 nm Ti and 100 nm Au were evaporated onto the backside surface to provide a homogeneous and conductive interface. Highly *n*-doped (As) and *p*-doped (B) Si substrates (Siegert Wafer, Germany) with a <100> crystal orientation and a resistivity of ~0.005 Ω cm were used for KPFM.

**Optical lithography.** Lithographic lift-off processing was combined with low-temperature plasma-assisted ALD to define a micropattern of $AlO_x$ on the frontside of Si substrates. A resist double layer consisting of poly-dimethyl glutarimide (PMGI SF3, Microchem) and ma-P 1205 (micro resist technology) was patterned with a maskless aligner (Heidelberg, MLA100) and developed in tetramethylammonium hydroxide (TMAH, micro resist technology). Lift-off was performed by ultrasonication (10 min, 40 °C) in n-methyl-2-pyrrolidone (NMP, micro resist technology).

**Atomic layer deposition.** Plasma-assisted ALD was carried out in a hot-wall reactor (Fiji G2, Veeco) at either 200 °C or 70 °C using Ar (99.9999%, Linde) as the carrier and purge gas. $AlO_x$ was formed on the Si substrate following sequential exposure to trimethylaluminum (TMA, 99.999%, Strem) and remote hydrogen plasma that was generated with a sapphire-based inductively coupled plasma source. The TMA half-cycle was carried out at a background pressure of 0.09 Torr and at a peak pressure of 0.16 Torr, after which the second half-cycle comprised two consecutive $H_2$ plasma exposures at 100 W at a background pressure of 0.01 Torr for a duration of 2 s (3s). In a separate process, TMA and deionized water (HPLC grade, Sigma Aldrich) were used for conventional thermal ALD of alumina at 200 °C. For all processes, purging between half cycles was performed using flowing Ar (110 sscm, 0.09 Torr) for 15 s. The purge time was set to 15 s or 60 s for ALD processes carried out at a reactor temperature of 200 °C or 70 °C, respectively. For all substrates, *in situ* cleaning of Si surfaces *via* remote hydrogen plasma exposure (100 W, 20 sscm, 0.3 Torr) was performed to remove adventitious organic contamination and activate the surface.[28]

**Spectroscopic ellipsometry.** Changes in the adlayer and film thickness on the Si substrates were monitored *in situ* during ALD with a sampling time of ~3 s using a spectroscopic ellipsometer (M-2000, J. A. Woollam). The light of a Xenon lamp (Hamamatsu, L2174-01) was focused to a spot area of ~5×8 $mm^2$ onto the substrate surface. The incoming and reflected light passed through fused silica windows (Lesker, VPZL-275DU) arranged in a fixed angle (67°) geometry. We obtained the thickness of the $SiO_2$ layers and the ALD $AlO_x$ film by analyzing the ellipsometry spectra with established optical models for Si and SiO$_2$ and by using a Cauchy model for the AlO$_x$ layer. Additional experimental details are provided in the **Supporting Information S1**.

**X-ray photoelectron spectroscopy.** XPS spectra were acquired in the hybrid lens mode at a pass energy of 10 eV and a take-off angle of 0° with a Kratos Axis Supra setup equipped with a monochromatic Al Kα X-ray source (photon energy = 1486.7 eV) operated with an emission current of 15 mA. The beam area was set to ~2×1 mm$^2$ using the slot collimation mode. The binding energy was calibrated with *in situ* sputter-cleaned Ag, Au, and Cu standard samples. Thereby, the kinetic energies of the Ag 3d (1118.51 eV), Au 4f (1402.73 eV) and Cu 2p (554.07 eV) core levels were referenced with an accuracy of 25 meV to the known peak values. Charge neutralization was not required as no binding energy shifts indicative of (differential) charging were observed for the doped Si substrates.

In addition, XPS measurements were performed with a custom-built system utilizing components from SPECS Surface Nano Analysis GmbH (Germany), a dual (Al and Mg) non-monochromatic X-ray tube anode, a PHOIBOS 100 concentric hemispherical analyzer, and an MCD-5 detector. Measurements were performed at a base pressure below $7\times10^{-9}$ mbar. The Al X-ray source (Al K$_\alpha$ = 1486.6 eV) was operated at an emission current of 20 mA with a voltage of 12.5 kV. XPS spectra were recorded with a pass energy of 20 eV or 10 eV in normal emission (90° ± 2° relative to the surface) in medium area mode (circular electron collection area of 2 mm diameter). Measurement parameters are transmission optimized, resulting in an acceptance angle up to ± 7°. With these settings an instrumental resolution of 1.2 eV was determined by measuring the full width at half maximum (FWHM) of a Au 4f$_{7/2}$ XPS spectrum recorded on a 100 nm Au layer evaporated on Si(100) substrate. The binding energy of was calibrated with Ag and Au standard samples. XPS spectra were recorded with SpecsLab (SPECS Surface Nano Analysis GmbH, ver. 2.85).

**Quadrupole mass spectrometry.** To detect gas phase species *in situ* during ALD, mass spectrometry experiments were conducted with a Hiden mass spectrometer (HPR-30), equipped with two detectors, a Faraday cup, and a secondary electron multiplier, the latter of which was used for all experiments. The tube (diameter 5 cm) connecting the mass spectrometer to the reactor was heated to above 120 °C. An orifice with an opening of 0.05 mm was used.

**Surface photovoltage measurements.** SPV measurements were carried out in ambient air at room temperature using a commercial setup (KP, KP Technology) in a closed faraday cage. The Si substrate was illuminated with a focused light-emitting diode with a wavelength of 880 nm (M340L4, Thorlabs) and an intensity of ~35 mW/cm$^2$. A piezoelectrically driven gold probe with a tip diameter of 50 μm and a work function of ~4.9 eV was used as the measurement electrode. The time resolution of the CPD measurements was 1.0 s.

**Scanning photoelectron microscopy.** Synchrotron-based SPEM measurements were performed at the ESCA microscopy beamline at the Elettra synchrotron (Trieste, Italy). Technical details of the SPEM setup are described in a previous report.[47] In brief, the X-ray photon energy was set to 638 eV with a (Gaussian) X-ray focal radius of ~150 nm and pass energy (PE) of 20 eV, resulting in an instrumental broadening of 0.35 eV determined by the Au 4f core level emission from an internal gold standard. Maps were acquired by using a multichannel delay line detector, allowing the acquisition of a spectrum in snap-shot mode at every pixel of the map.

**Scanning electron microscopy.** The SEM images were acquired with an NVision 40 FIB-SEM from Carl Zeiss using the in-lens secondary electron detector and an electron beam acceleration voltage of 2.0 kV.

**Atomic force microscopy and Kelvin probe force microscopy.** AFM and KPFM measurements were carried out with a Bruker Multimode V microscope (Billerica, MA, USA) in ambient using platinum/iridium (95/5) coated AFM probes (PPP EFM, Nanosensors) with a nominal tip radius of 20 nm, typical resonance frequency of 75 kHz and force constant of 2.8 N/m. Height images (12×12 μm$^2$) were acquired at a scan rate of 0.3 Hz with 512 point sampling. High-resolution images were obtained with a 2×2 μm$^2$ scan with 512 point sampling using an uncoated Si tip with ~10 nm tip radius.

## ASSOCIATED CONTENT

**Supporting Information**. The following files are available free of charge.

Additional details on data modelling and analysis, as well as SE, QMS, XPS, AFM and SPV measurements. Supporting static water contact angle measurements. (PDF)

## AUTHOR INFORMATION

**Corresponding Author**

*sharp@wsi.tum.de, *alex.henning@wsi.tum.de

**Present Addresses**

[+]Department of Chemistry and Bar-Ilan Institute for Nanotechnology & Advanced Materials, Bar-Ilan University, Anna and Max Webb Street, Ramat Gan 5290002, Israel

[#]Helmholtz-Zentrum Berlin für Materialien and Energie GmbH, BESSY II, Albert-Einstein-Straβe 15, 12489 Berlin, Germany

[%]Fritz-Haber-Institut der Max-Planck-Gesellschaft, Dept. Inorganic Chemistry, Faradayweg 4-6, 14195 Berlin, Germany

**Author Contributions**

All authors have given approval to the final version of the manuscript.

**Funding Sources**

This work was supported by the Deutsche Forschungsgemeinschaft (DFG, German Research Foundation) under Germany´s Excellence Strategy – EXC 2089/1 – 390776260 and through the TUM International Graduate School of Science and Engineering (IGSSE), as well as the Federal

Ministry of Education and Research (BMBF, Germany) project number 033RC021B within the CO2-WIN initiative. AH acknowledges funding from the European Union's Horizon 2020 research and innovation programme under the Marie Skłodowska-Curie grant agreement No 841556.

# Supporting Information

# Spatially-Modulated Silicon Interface Energetics *via* Hydrogen Plasma-Assisted Atomic Layer Deposition of Ultrathin Alumina

*Alex Henning[§, *], Johannes D. Bartl[§, ‡, +], Lukas Wolz[§], Maximilian Christis[§], Felix Rauh[§], Michele Bissolo[§], Theresa Grünleitner[§], Johanna Eichhorn[§], Patrick Zeller[†, #, %], Matteo Amati[†], Luca Gregoratti[†], Jonathan J. Finley[§], Bernhard Rieger[‡], Martin Stutzmann[§], Ian D. Sharp[§, *]*

[§]Walter Schottky Institute and Physics Department, Technical University of Munich, 85748 Garching, Germany

[‡] Department of Chemistry, WACKER-Chair for Macromolecular Chemistry, Technische Universität München, Lichtenbergstraße 4, 85747 Garching bei München, Germany

[†] Elettra-Sincrotrone Trieste SCpA, SS14-Km163.5 in Area Science Park, 34149, Trieste, Italy.

*sharp@wsi.tum.de, *alex.henning@wsi.tum.de

## S1. Spectroscopic ellipsometry (SE) of Si substrates during atomic layer deposition (ALD)

We considered a three-layer model structure to analyze the SE spectra of an amorphous aluminum oxide ($AlO_x$) film on a Si substrate covered with silicon dioxide ($SiO_2$). To obtain the thickness of the $SiO_2$ layer, we first characterized a $Si/SiO_2$ substrate by SE. A temperature-dependent library model and the Sellmeier model were used to describe the optical properties of the silicon substrate and the $SiO_2$ layer, respectively. Because alumina is optically transparent in the measured wavelength range (210 nm – 1800 nm), it is valid to use the Cauchy dispersion model, $n(\lambda) = A + \frac{B}{\lambda^2} + \frac{C}{\lambda^4}$, where n is the refractive index, λ is the wavelength, and A, B and C are material-dependent coefficients, to track thickness changes during the formation of $AlO_x$. We determined the parameters for ALD alumina (A = 1.751, B = 6.320×10$^{-3}$ and C = -1.052×10$^{-4}$) from fitting the SE data of a thermally grown $AlO_x$ film of known thickness (10 nm) on a silicon substrate, resulting in a refractive index of ~1.8 for $AlO_x$ within the measured wavelength range.

Before deposition of alumina, we removed the adsorbate (water and hydrocarbons) of the solvent-cleaned Si substrates with a remote $H_2$ plasma pretreatment. As evident from the evolution of the film thickness, obtained from *in situ* SE (Fig. S1, black curve), the initial $H_2$ plasma exposure reduces the adlayer film thickness by typically 2-4 Å. In this way, we omit wet-chemical treatments (*e.g.*, RCA processing, hydrofluoric or sulfuric acid treatment) that are typically used for Si surface cleaning or activation, and facilitate process reproducibility. We note that, while oxygen plasma during ALD of $AlO_x$ may lead to unwanted surface oxidation and defect generation,[1] atomic hydrogen and moderate doses of hydrogen plasma have a minor effect on interface state generation and can even passivate surface defects in Si.[2,3]

In a control SE experiment with argon (Ar) plasma instead of hydrogen ($H_2$) plasma, we observed a substantially larger growth rate (Fig. S1, blue curve), which is associated with an incorporation of carbon impurities, as suggested by the strong decrease in the adlayer film thickness during subsequent oxygen plasma treatment. In contrast to argon plasma, the remotely generated $H_2$ plasma reacts with the adsorbed TMA to produce methane and potentially reduces the silicon oxide *via* formation of silane ($SiH_4$) gas, although the latter could not be verified by *in situ* mass spectrometry due to lack of sensitivity.

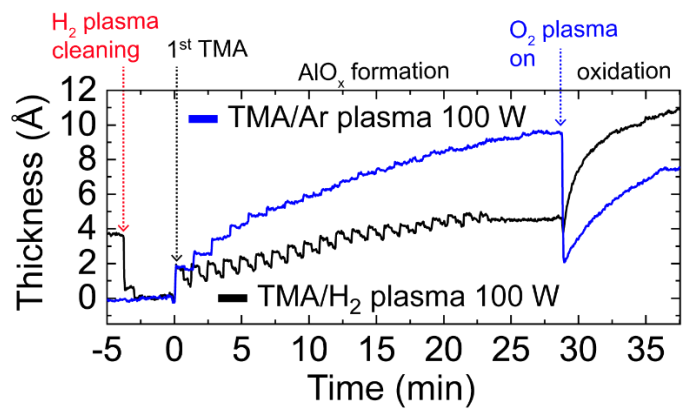

**Figure S1.** Thickness of surface oxide layers as a function of time plot obtained from *in-situ* SE on Si substrates with a native SiO$_2$ layer during 20 cycles of the TMA/H$_2$ (black line) and TMA/Ar plasma (blue line) process at 200 °C and 100 W plasma power. The oxygen plasma treatment was carried out at 100 W plasma power and a 20 sscm flow rate and 0.04 Torr background pressure.

## S2. *In situ* quadrupole mass spectrometry (QMS) during ALD

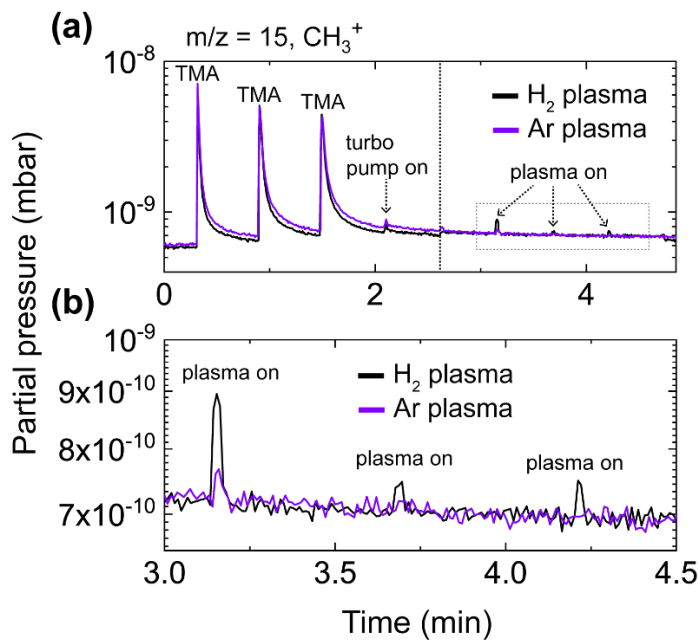

**Figure S2.** (a) Tracking of the methane vapor during the introduction of TMA and $H_2$ plasma (black line) or Ar plasma (purple line) into the reactor chamber by *in situ* mass spectrometry. The $CH_3^+$ ions with a mass-to-charge (m/z) ratio of 15 stem from $Al(CH_3)_3$ (i.e., TMA) or methane ($CH_4$). (b) The zoom in into the region marked in (a) highlights $CH_3^+$ species detected during the plasma steps of the TMA/$H_2$ plasma and TMA/Ar plasma processes.

## S3. X-ray photoelectron spectroscopy (XPS) measurements of Si substrates

The absence of stoichiometric $SiO_2$ component and concurrent appearance of sub-stoichiometric $SiO_x$ (Fig. 2b) and an alumina component (Al 2p) following 20 cycles $TMA/H_2$ plasma (Fig. 2b, inset) hint to the formation of a mixed phase (*i.e.*, Si-O-Al bonding), although the creation of oxygen-deficient $SiO_x$ with oxygen vacancies cannot be excluded from the XPS data. We note that we found no evidence by XPS (Figs. 2 and S3) for the formation of an $AlSi_xO_y$ interfacial layer that has been previously reported for Si substrates following ALD of $AlO_x$ and post-deposition annealing.[4]

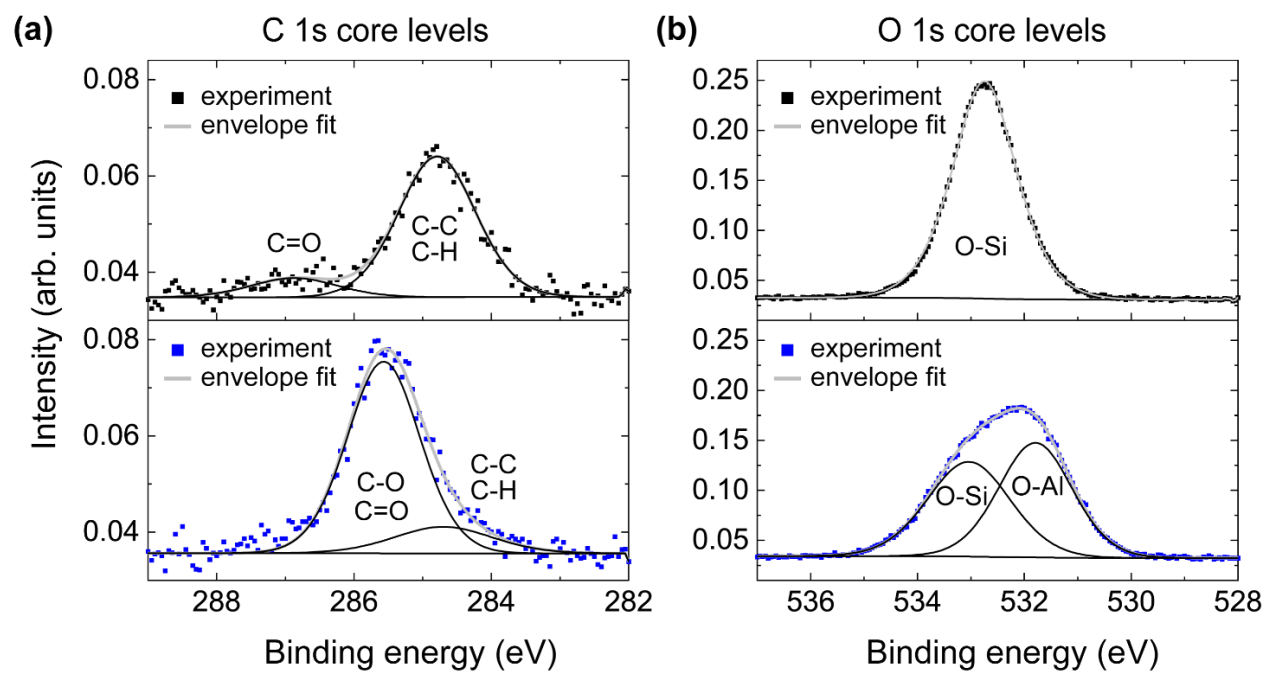

**Figure S3.** (a) Carbon 1s (C 1s) and (b) oxygen 1s (O 1s) core level spectra of solvent-cleaned bare *p*-doped Si (top, black squares) and *p*-Si substrates after exposure to 20 cycles (bottom, blue squares).

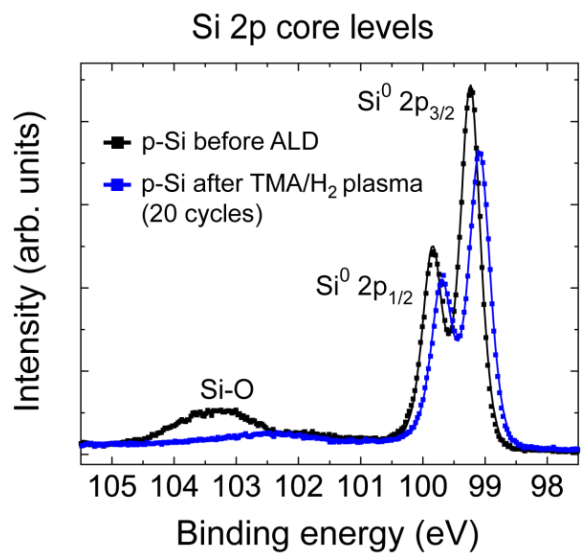

**Figure S4.** Silicon 2p core level spectra of silicon substrates without (black) and with (blue) an alumina layer deposited by ALD following 20 cycles of TMA and remote $H_2$ plasma. The intensity of detected photoelectrons is not normalized in this plot.

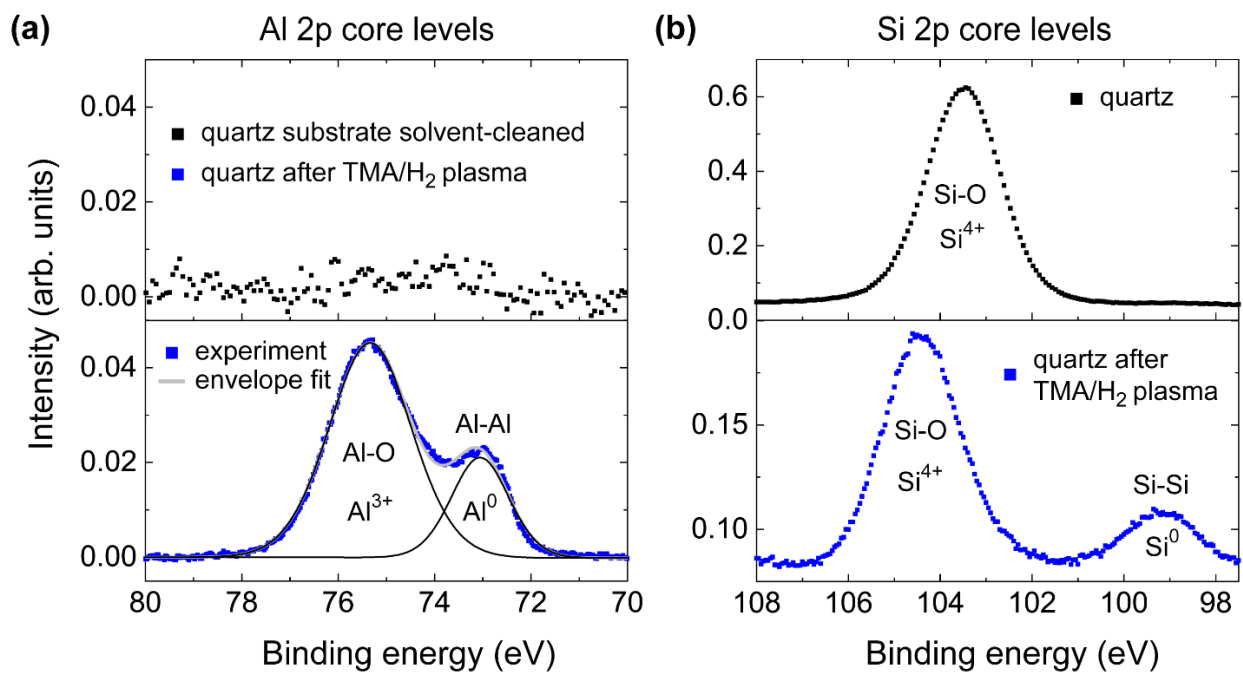

**Figure S5.** X-ray photoelectron spectra of the (a) Al 2p and (b) Si 2p core levels of a representative solvent-cleaned quartz substrate (top row, black squares) and quartz substrates after TMA/$H_2$

plasma (200 cycle, 300 W) treatment (bottom row, blue squares). The two peaks in the Al 2p spectrum of the ALD-processed quartz substrates appear at binding energies of 75.5 eV and 73 eV, which are characteristic of aluminum oxide and elemental aluminum, respectively. Concurrently, a spectral signature of elemental silicon at a binding energy around 99 eV emerges in the photoelectron spectrum of Si 2p.

## S4. Overview of SiO$_2$ thicknesses and static water contact angles of Si substrates

Static contact angle (SCA) measurements were performed with the contact angle system OCA 15Pro (DataPhysics Instruments GmbH, Baden-Wuerttemberg, Germany) on Silicon substrates under ambient conditions (27.3 °C, 32.7 % relative humidity). Data were acquired and evaluated with the basic module SCA 20 - contact angle (DataPhysics Instruments GmbH, Baden-Wuerttemberg, Germany, v. 2.0). To estimate an average Young-LaPlace contact angle, 1 µl of deionized H$_2$O (18.2 MΩ·cm at 25 °C, Merck Millipore) was dispensed with a rate of 1 µl/s from a high-precision syringe (Hamilton, DS 500/GT, gas-tight, 500 µl) on the sample surface and after ~3 s (reaching equilibrium) the side profile of the droplet was taken for further processing. SCAs from at least three different spots were determined to calculate a standard deviation (Table S1).

**Table S1.** Static water contact angle measurement on silicon substrates before and after treatment (TMA and remote H$_2$ plasma at 200 °C) in a hot wall ALD reactor. The thicknesses of the (native) silicon oxide layers were determined by spectroscopic ellipsometry and XPS.

| Process details | Surface oxide | $t_{SiO2}$ by SE | $t_{SiO2}$ by XPS | water CA / ° |
|---|---|---|---|---|
| Solvent-cleaned | Si/SiO$_2$ | 16 Å | 17 Å | 64.0 ± 0.5 |
| 3 cycles remote H$_2$ plasma treatment | Si/SiO$_2$ | 16 Å | / | 34.2 ± 0.6 |
| 20 cycles TMA/H$_2$ plasma (100 W) | Si/SiO$_2$/AlO$_x$ | / | 8 Å | 24.7 ± 0.7 |
| 100 cycles TMA/H$_2$ plasma (100 W) | Si/AlO$_x$ | / | 0 Å | / |
| 12 cycles TMA/H$_2$O | Si/SiO$_2$/AlO$_x$ | / | 18 Å | 10.0 ± 1.0 |

## S5. Atomic force microscopy and Kelvin probe force microscopy of dielectric patterns

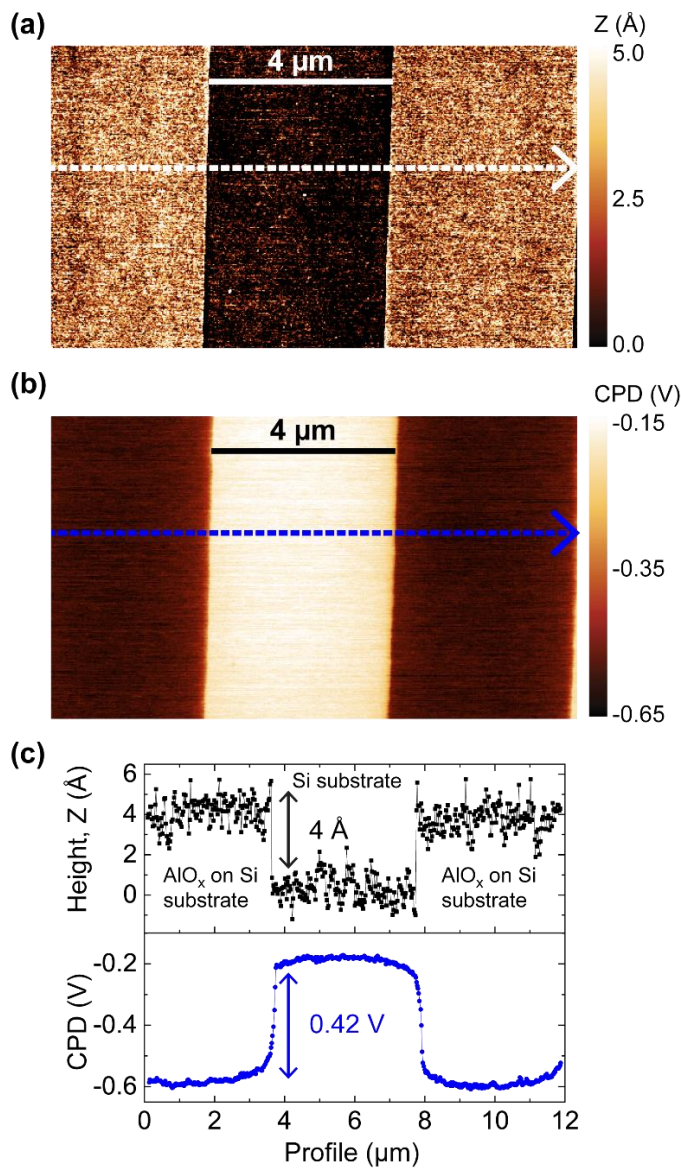

**Figure S6.** (a) Intermittent contact mode AFM topography and (b) frequency modulation KPFM image of the AlO$_x$ pattern on *p*-doped Si substrate. (c) Topography and CPD profiles were taken along the lines, as indicated in the images of (a) and (b). A step height of 4 Å exists across the lateral interface between AlO$_x$ and SiO$_2$ terminal surfaces, and a surface potential step of 0.42 V arises due to fixed charge introduced by the TMA/H$_2$ ALD process (20 cycles).

## S6. Surface photovoltage (SPV) measurements and calculation of induced surface charge

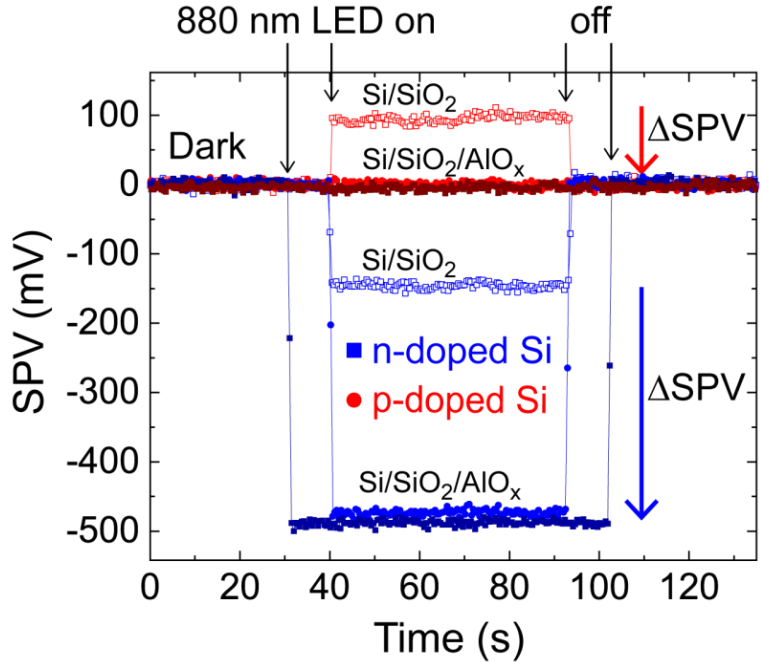

**Figure S7.** Plots of the surface photovoltage as a function of time upon switching the 880 nm illumination source on and off for *p*-doped Si (dark red) and *n*-doped Si (dark blue) before (open circles) and after (filled circles) processing by 12 cycles of TMA and H$_2$O, and the plots of the SPV as a function of time for *p*-doped Si (red) and *n*-doped Si (blue) before (open circles) and after (filled circles) processing by 20 cycles of TMA and remote H$_2$ plasma. The vertical arrows indicated the magnitude of SPV change induced by the ALD AlO$_x$ processes.

We used established theory for semiconductor interfaces[5] to estimate the surface charge density introduced by the ALD AlO$_x$ layer on the Si surface based on the measured change in the surface photovoltage (ΔSPV) (Fig. S6). For the *n*-doped Si, the change in the SPV approximately equals the change in the surface band bending.

Because of charge neutrality, the negative surface charge per unit area, $Q_{AlOx}$, introduced by the sub-nanometer thin AlO$_x$ coating at the AlO$_x$/SiO$_2$ interface, is compensated by the positive charges, $Q_{Si}$, in the Si surface space charge region:

$$Q_{it} = -Q_{Si} \qquad (1).$$

We calculated the charge per unit area that is introduced by the alumina coating with Equation (2) that was obtained by solving the Poisson equation for a one-dimensional semiconductor model (Fig. 5b of the main manuscript) and applying Gauss' theorem:

$$Q_{Si} = -\sqrt{2\varepsilon_{Si} kT N_D} \left[ \left( e^{-qV_S/kT} + \frac{qV_S}{kT} - 1 \right) + \frac{n_i^2}{N_D^2} \left( e^{qV_S/kT} - \frac{qV_S}{kT} - 1 \right) \right]^{1/2} \qquad (2)$$

where $\varepsilon_{Si}$ is the relative permittivity of Si, $k$ is the Boltzmann constant, $T$ is the temperature, $N_D$ is the donor concentration, $q$ is the elementary charge, $V_S$ (≈ ΔSPV) is the surface band bending and $n_i$ is the Si intrinsic carrier concentration. We note that Equation (2) is written in the form applicable for *n*-type semiconductors, for which electrons are the majority carriers and holes are the minority carriers.